\documentclass[%
 reprint,
 amsmath,amssymb,
 aps,
]{revtex4-2}
\usepackage[allow-number-unit-breaks]{siunitx}
\usepackage{graphicx}
\usepackage{dcolumn}
\usepackage{bm}
\usepackage[utf8]{inputenc}
\usepackage[T1]{fontenc} 
\usepackage{wasysym} 
\usepackage{subcaption}
\usepackage{makecell} 
\usepackage{array} 
\newcolumntype{P}[1]{>{\centering\arraybackslash}p{#1}}
\newcolumntype{M}[1]{>{\centering\arraybackslash}m{#1}}
\usepackage[version=4]{mhchem} 
\usepackage{placeins}
\usepackage{siunitx} 

\newcommand{\mm}[1]{\SI{#1}{\micro \meter}} 

\newcommand{\mb}[1]{\mathbf{#1}}

\newcommand{\fref}[1]{Fig.~\ref{fig:#1}}

\newcommand{\sdfrac}[2]{\mbox{\footnotesize$\displaystyle\frac{#1}{#2}$}}
\newcommand{\nexp}{$n_{\mathrm{exp}}$}

\begin{document}

\preprint{APS/123-QED} 

\title{Tilted Dirac cones and asymmetric conical diffraction \\
in photonic Lieb-kagome lattices}

\author{Jean-Philippe Lang}
\author{Haissam Hanafi}%
 \email{haissam.hanafi@uni-muenster.de}
\author{J{\"o}rg Imbrock}%
\author{Cornelia Denz}%
\affiliation{%
Institute of Applied Physics, University of M{\"u}nster, 48149 M{\"u}nster, Germany
}%




\date{\today}

\begin{abstract}
The Lieb lattice and the kagome lattice, which are both well known for their Dirac cones and flat bands, can be continuously converted into each other by a shearing transformation. During this transformation, the flat band is destroyed, but the Dirac cones remain and become tilted, with types I, II, and III occurring for different parameters. In this work, we first study these tilted Dirac cones using a tight-binding model, revealing how they can be engineered into the different types. We then demonstrate conical diffraction in a photonic lattice realization of the Lieb-kagome lattice using split-step beam propagation simulations, obtaining evidence of the presence of Dirac cones tilted in different directions. Finally, we performed experiments with photonic lattices laser-written in fused silica (\ce{SiO2}) to validate the results of the simulations. These studies advance the understanding of the Lieb-kagome lattice and tilted Dirac cones in general and provide a basis for further research into this interesting tunable lattice system. 

\end{abstract}

\maketitle




\section{Introduction\label{sec:Introduction}} 

A Dirac cone is an intersection of (energy-)bands in a single point, the Dirac point, surrounded by linear dispersion, thus forming cones in the band structure. In particular, since the discovery of graphene \cite{Novoselov2004}, in which Dirac cones exist, they have been the focus of both theoretical and experimental research interest \cite{CastroNeto2009, Wehling2014, Yang2016, Wang2015}. Aside from the regular Dirac cones found, among others, in graphene, there exist also more exotic Dirac cones, such as higher-order conical intersections \cite{Leykam2016} or tilted Dirac cones \cite{Milicevic2019, Kawarabayashi2011,  Cheng2017}. Tilted Dirac cones can be classified according to their degree of tilting into weakly tilted (type I), strongly tilted (type II), and critically tilted (type III) cones \cite{Milicevic2019}. Type III cones in particular remain challenging to observe in solid-state physics, although some candidates for a realization have recently been suggested \cite{Huang2018, Sims2021}. Since the origin of Dirac cones lies in lattice symmetries, however, they are fundamental phenomena, which can occur in any periodic system. One such system that is of particular interest due to its versatility is a photonic lattice, which consists of a periodic arrangement of weakly coupled single-mode waveguides \cite{Szameit2010}. Using fabrication techniques such as direct laser writing arbitrary lattice geometries in different host materials can be realized \cite{imbrock2018local, hanafi2019polycrystalline, imbrock2022thermally}.By varying the shape of the constituting waveguides in a photonic lattice, one can simulate the influence of electrical fields, realizing phenomena such as Floquet topological insulators \cite{Rechtsman2013}, Bloch-Zener oscillations \cite{dreisow2009bloch} or dynamic localization \cite{longhi2006observation}. Recently, even experimental evidence of type III Dirac cones has been found in a photonic lattice system \cite{Milicevic2019}. \\Spatial light evolution in a photonic lattice is described by the paraxial wave equation, which is mathematically equivalent to a time-dependent Schrödinger equation. Therefore, photonic lattices can be used as model systems for the time evolution of the electron-wave function in 2D materials. In photonic lattices, the presence of Dirac cones leads to a phenomenon known as conical diffraction, in which a ring of light with constant thickness and linearly growing diameter is observed. Such conical diffraction has been shown in experiments for regular Dirac cones \cite{Peleg2007, Diebel2016}. Conical diffraction from tilted Dirac cones was recently demonstrated in simulations \cite{Zhong2019a}. \\
The Lieb lattice and the kagome lattice are two types of artificial lattices, which have long been studied theoretically, because they feature both Dirac cones and completely dispersionless flat bands. They can be easily realized as photonic lattices, and as such have been studied in the context of localized (flat band) states \cite{Vicencio2015,zong2016observation,hanafi2022APL}, conical diffraction \cite{Leykam2012, Diebel2016, Liu2021}, and topological insulators\cite{Zhong2019}. The Lieb and the kagome lattice are related by a shearing transformation and can be continuously transformed into each other. Such a Lieb-kagome model was, to our knowledge, first proposed in 2011 \cite{Asano2011}, and recently Jiang et al. \cite{Jiang2019} published a study on this lattice system focusing on the topological effects caused by breaking the time-reversal symmetry. Soon after, Lim et al. \cite{Lim2020} released a detailed theoretical study on the splitting of Dirac cones during the transition from Lieb to the kagome lattice. So far none of the works on the Lieb-kagome lattice contain simulations or experiments, that could verify the numerical results. Furthermore, the variable tilting of the Dirac cones of this lattice system has not yet been studied systematically. Doing so would reveal ways to engineer not only the rather common tilted type I Dirac cones, but also the more rare type II and III cones. Type II Dirac cones can be used in photonic lattices to study several interesting phenomena, such as Klein tunneling \cite{Jin2020}, topological valley Hall states \cite{Zhong2021}, and more. Type III cones are of interest as a model system for a black hole event horizon \cite{Huang2018, Chen2020}. To fill these gaps, in this work, we first study the Lieb-kagome lattice with tight-binding calculations, focusing on the variable tilting of its Dirac cones, showing how it can be tailored to realize type I, II, and III Dirac cones. After that we demonstrate asymmetric conical diffraction in photonic Lieb-kagome lattices, resulting from differently tilted Dirac cones, by performing simulations based on the split-step algorithm \cite{sharma2004new}. Lastly, we perform experiments in laser-written photonic lattices, testing the results of the previous simulations. 

\begin{figure*}[htbp]
	\begin{subfigure}[b]{0.25\textwidth}
		\includegraphics[width=\textwidth]{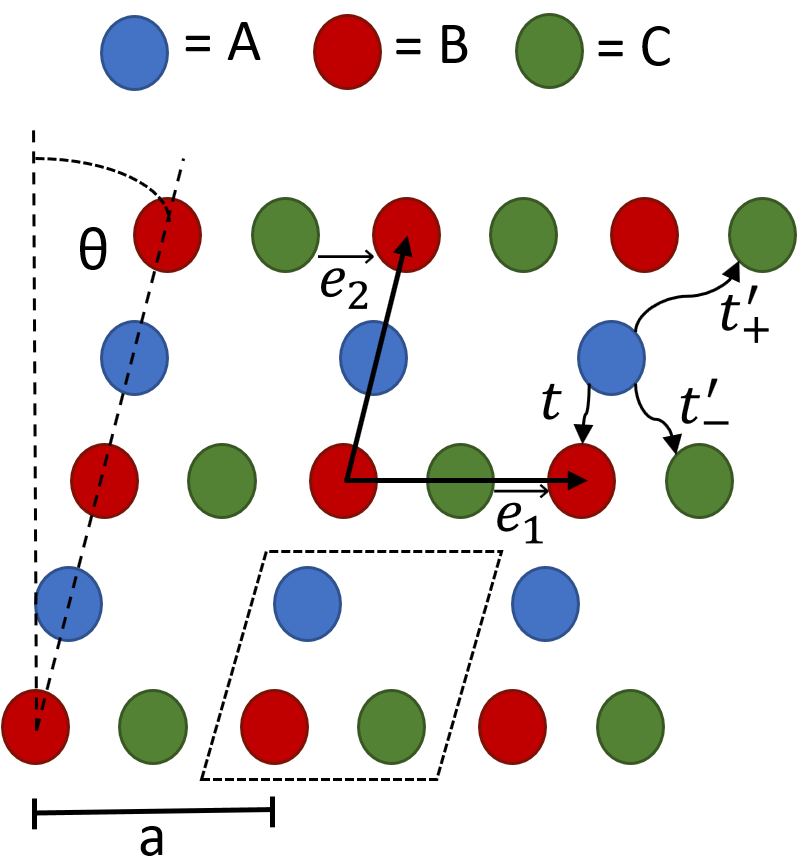}
		\caption{}
		\label{fig:Lk1}
	\end{subfigure}
	~
	\begin{subfigure}[b]{0.4\textwidth}
		\includegraphics[width=\textwidth]{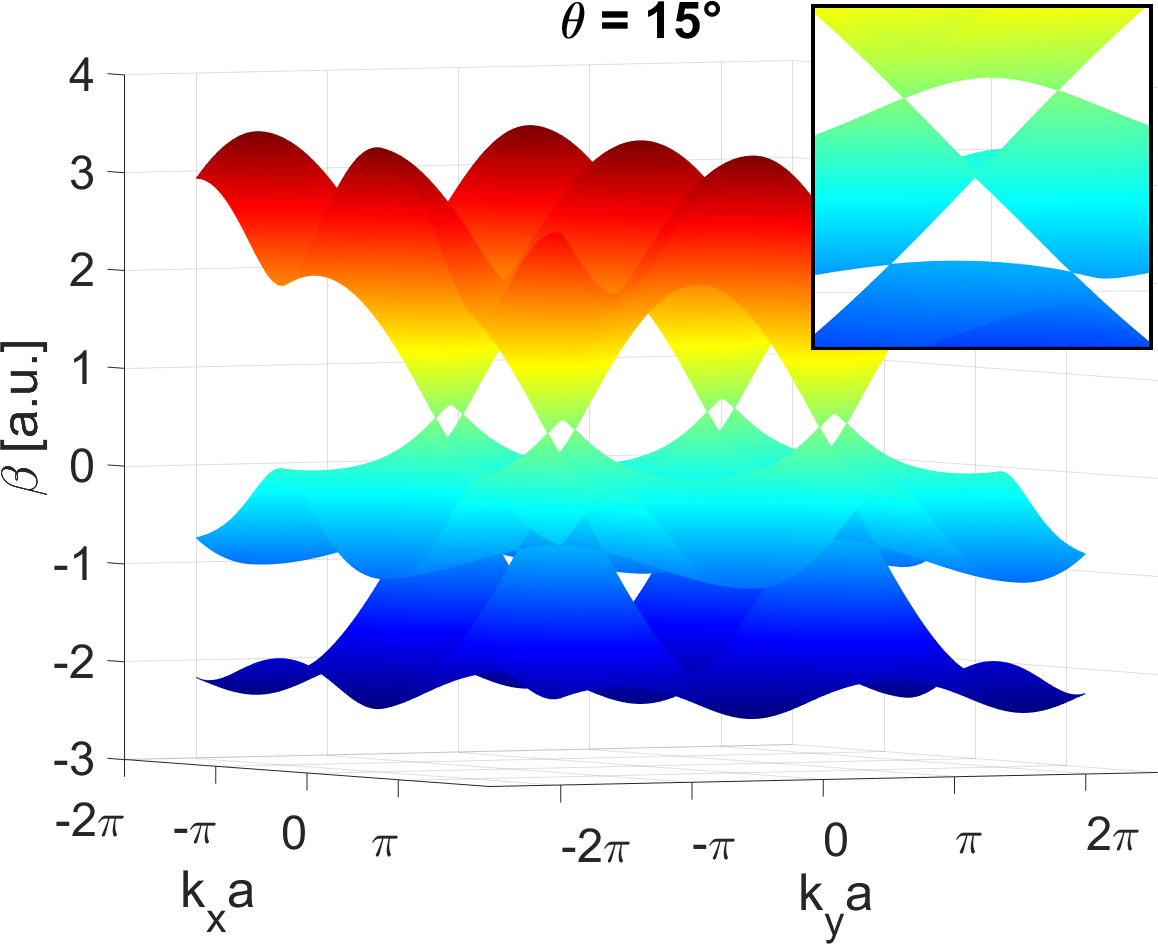}
		\caption{}
		\label{fig:Lk2}
	\end{subfigure}
	~
	\begin{subfigure}[b]{0.3\textwidth}
		\includegraphics[width=\textwidth]{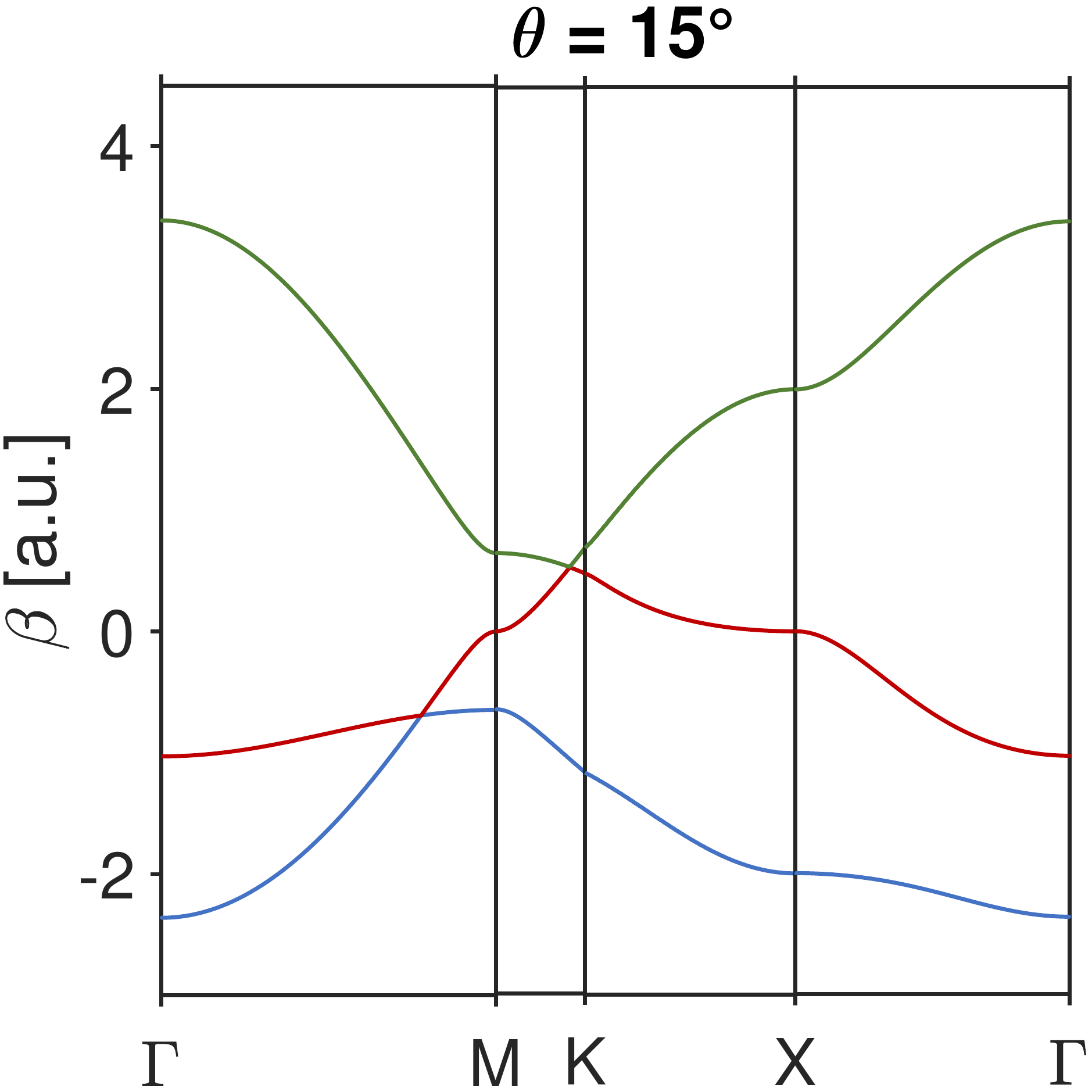}
		\caption{}
		\label{fig:Lk3}
	\end{subfigure}\\[20pt]
	
	\begin{subfigure}[b]{0.3\textwidth}
		\includegraphics[width=\textwidth]{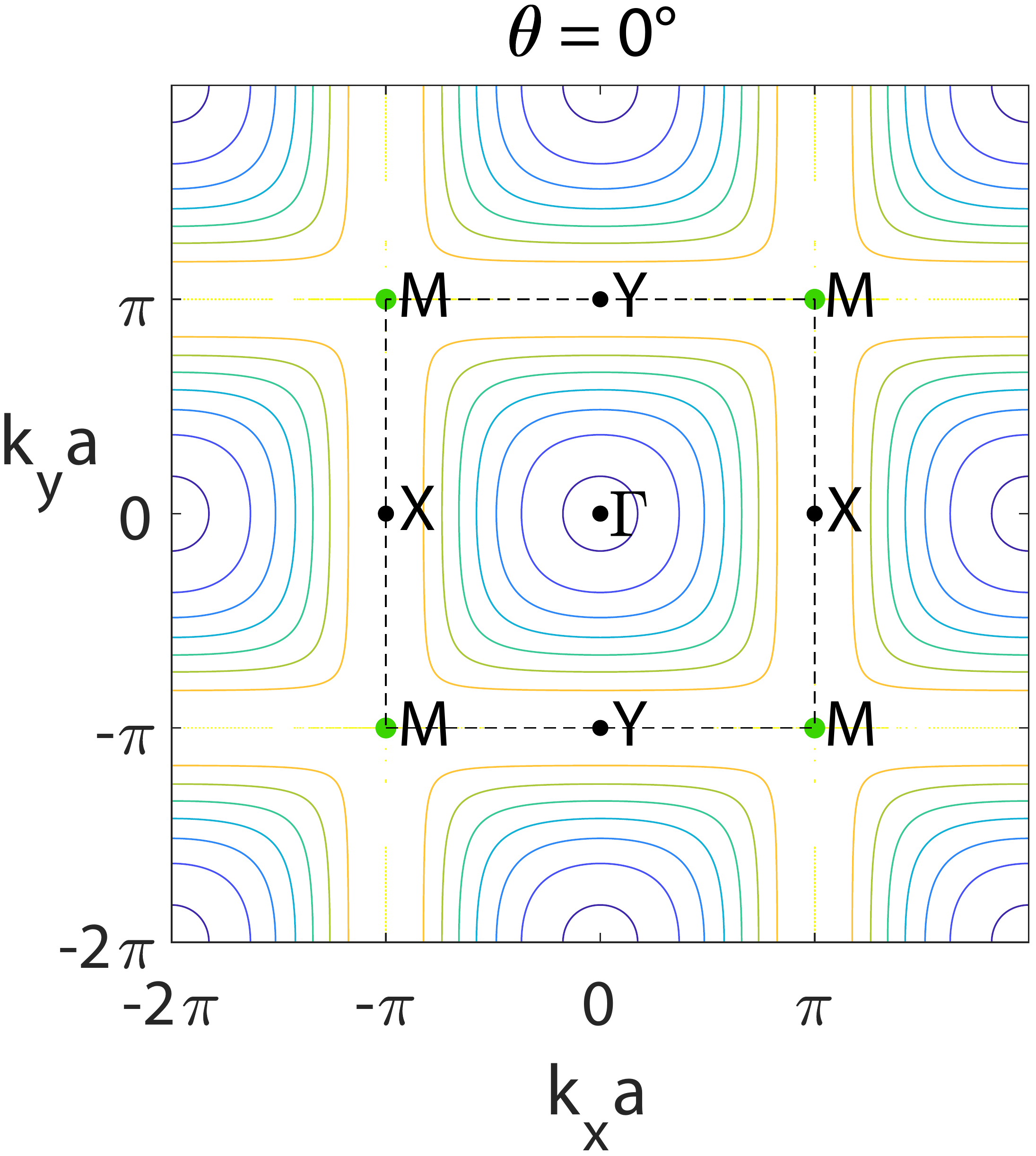}
		\caption{}
		\label{fig:gr1}
	\end{subfigure}
	~
	\begin{subfigure}[b]{0.3\textwidth}
		\includegraphics[width=\textwidth]{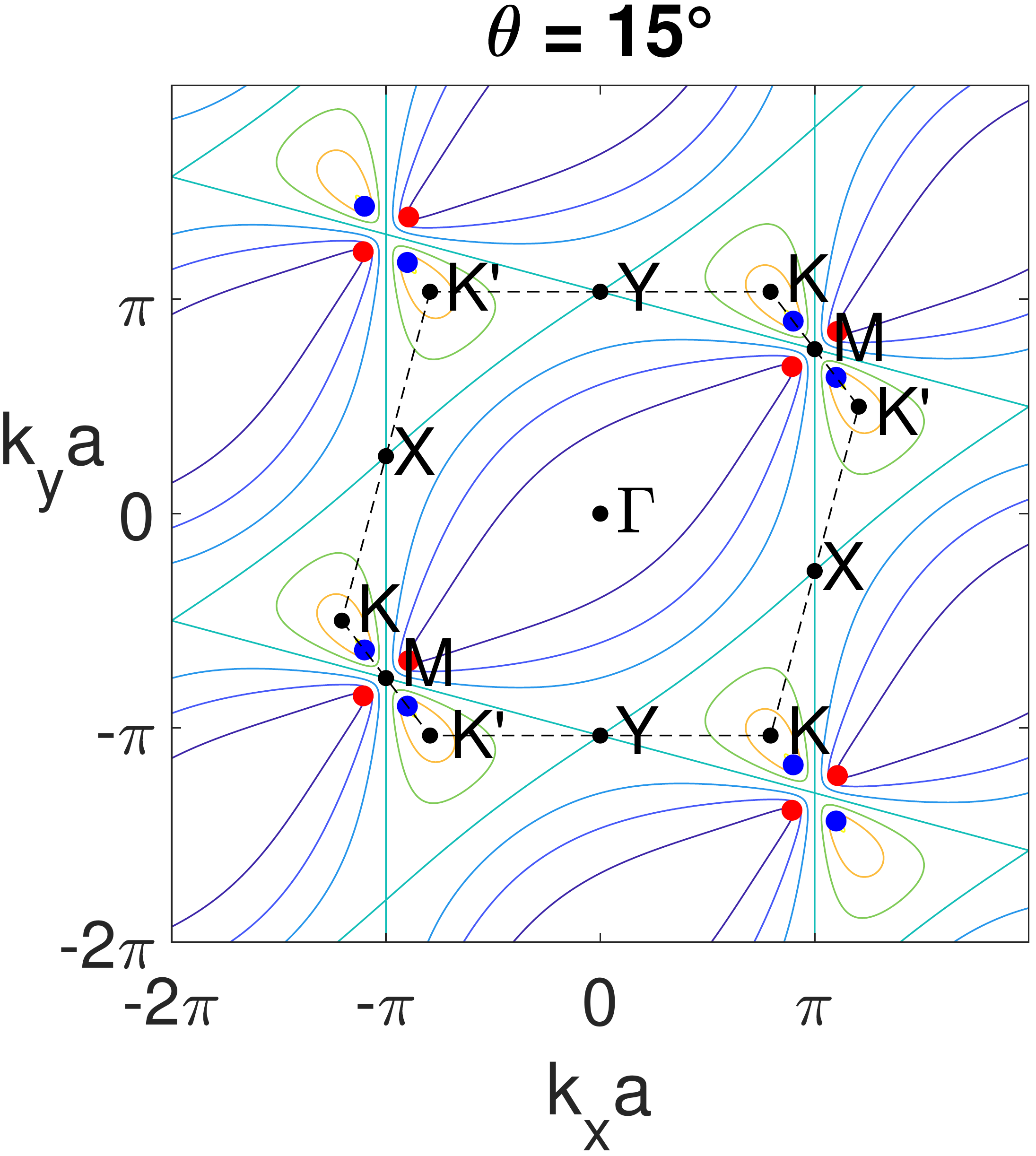}
		\caption{}
		\label{fig:gr2}
	\end{subfigure}
	~
	\begin{subfigure}[b]{0.3\textwidth}
		\includegraphics[width=\textwidth]{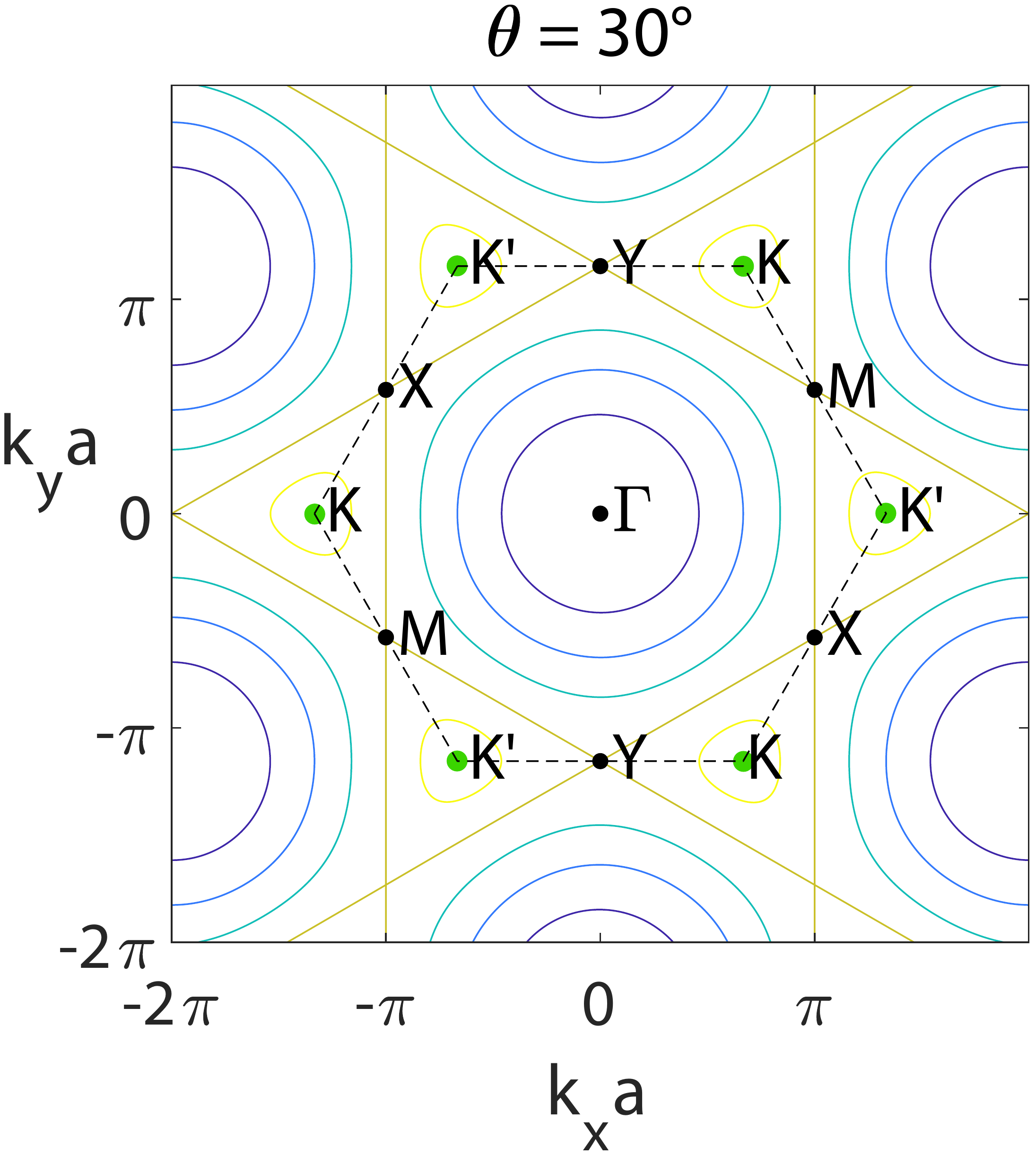}
		\caption{}
		\label{fig:gr3}
	\end{subfigure}
	\caption{The Lieb-kagome lattice \textbf{(a)} sketch of the Lattice geometry. \textbf{(b)} Tight-binding band structure of a Lieb-kagome lattice with $\theta=\ang{15}$. The inset shows a close-up of four pairs of Dirac cones from a different angle. \textbf{(c)} projection of the band structure from (b) on lines between high-symmetry points. \textbf{(d - f)} Contour plots of the middle bands of Lieb-kagome lattices with different shearing angles $\theta$. The first Brillouin zones are indicated as dashed lines along with some high-symmetry points. Dirac points are indicated in green for regular Dirac cones, blue for the upper tilted Dirac cones, and red for the lower tilted Dirac cones.)}
	\label{fig:Lk-model}
\end{figure*}
 
\section{The Lieb-kagome lattice}\label{sec:lattice}

\subsection{Tight-binding model}\label{sec:model}

Although the Lieb and the kagome lattice belong to different symmetry groups, they share the structure of their unit cells, which consist of one corner site and two edge-centered sites \cite{Jiang2019}, making them interconvertable through a shearing transformation. Therefore, both lattices can be combined into a single unified model, which is sketched in \fref{Lk-model} (a). In this Lieb-kagome model we label the edge-center sites as A and C and the corner sites as B. The lattice constant $a$ is the distance between unit cells, making the nearest-neighbor distance $a/2$. The parameter that describes the transition between the Lieb and kagome lattice is the shearing angle labeled $\theta$. This angle can be varied between \ang{0}, at which point the model is equivalent to the Lieb lattice, and \ang{30}, where the model represents the kagome lattice. In our calculations, we consider nearest neighbor interaction, which occurs between lattice sites labeled A and B and between B and C, and next-nearest neighbor interaction between A and C sites. At $\theta=\ang{0}$ (Lieb lattice), every lattice site A has four C sites as next-nearest neighbors, all with the same distance. When $\theta$ increases, the distance to two of those C sites increases, while the distance to the other two decreases. At $\theta=\ang{30}$ (kagome lattice), an A lattice site has two C sites as nearest neighbors and two as next-nearest neighbors. To describe the Lieb-kagome lattice we therefore need three coupling constants: The nearest neighbor coupling constant $t$ and the next-nearest neighbor coupling constants $t'_{\pm}$ for the case of increasing ($t'_+$) or decreasing ($t'_-$) distances when $\theta$ is increased, as indicated in \fref{Lk-model} (a). To describe the distance dependence of the coupling constants, we adopt a model from \cite{Jiang2019} to describe the relative coupling strength $\gamma$:

\begin{equation}
	  \frac{t'}{t} \equiv \gamma  = \left[\exp{\left(\frac{d_{\mathrm{NN}}-d_{\mathrm{NNN}}}{d_{\mathrm{NN}}}\right)}\right]^{n_{\mathrm{exp}}}.
		\label{eq:gamma}
\end{equation}

Here, $d_{\mathrm{NN}}$ and $d_{\mathrm{NNN}}$ are the distances of nearest neighbor and next-nearest neighbor, respectively, while the exponent $n_{\mathrm{exp}}$ is a free parameter that determines how quickly the coupling decreases with distance. In our later performed simulations of conical diffraction we found $n_{\mathrm{exp}}=4$ to best describe our realistic photonic lattice system, as at that value clear ring patterns could be observed for all $\theta$ between \ang{0} and \ang{30}. Therefore, we used that value for all subsequent calculations, with the exception of the contour plots presented here, where we used $n_{\mathrm{exp}}=8$, because that value allows the movement of the Dirac cones to be seen more clearly. However, this value for \nexp~is still realistic for photonic lattices, particularly if the excitation uses a larger wavelength than we used in our simulations. Furthermore, the directions of movement of the Dirac cones and the conclusions drawn from it remain the same regardless of the specific value.\\ For the Lieb-kagome model shown in \fref{Lk-model} (a) some simple geometric considerations lead to

\begin{equation}
	 \frac{t'_\pm}{t} \equiv \gamma_\pm = \left[\exp{\left(1-\sqrt{2\pm 2\sin{\theta}}\right)}\right]^{n_{\mathrm{exp}}}.
\end{equation}  

After those deliberations, we can now calculate a momentum-space tight-binding Hamiltonian for the Lieb-kagome lattice:


\begin{equation}
	H(\mb{k})=2t
	\begin{pmatrix}
		0 & AB & AC \\
		AB & 0 & BC \\
		AC & BC & 0 
	\end{pmatrix},
\end{equation}

with the matrix elements

\begin{subequations}
\begin{alignat}{1}
	AB&=\cos{\left(\sdfrac{ak_x\sin{\theta}+ak_y\cos{\theta}}{2}\right)},\\[10pt]
	AC&=\gamma_-\cos{\left(\sdfrac{ak_x(1-\sin{\theta})-ak_y\cos{\theta}}{2}\right)}\nonumber \\[5pt] \quad&+\gamma_+\cos{\left(\sdfrac{ak_x(1+\sin{\theta})+ak_y\cos{\theta}}{2}\right)},\\[10pt]
	BC&=\cos{\left(\sdfrac{ak_x}{2}\right)}
\end{alignat}
\end{subequations}

corresponding to the interaction between the respective lattice sites. Diagonalizing this Hamiltonian yields three eigenvalues $\beta_n(k_x,k_y), n=1,2,3$. In a solid-state system, such eigenvalues represent the energy bands that describe the electron dynamics of the lattice. In a model system such as a photonic lattice, which we will consider later, they give us the diffraction relation $k_z(k_x,k_y)$.

\subsection{Band evolution}

\fref{Lk-model} (b) and (c) show the band structure and projection of it on lines between high-symmetry points of a Lieb-kagome lattice with $\theta=\ang{15}$, respectively. It can be seen that in transition states between Lieb and kagome lattice there is no longer a flat band, but there are several Dirac cones, which are clearly tilted. The Dirac points in transition states can be distinguished into those situated at positive $\beta$ and those at negative $\beta$, which we will call in the following upper and lower Dirac points/cones, respectively. \\The movement of Dirac points in dependence on $\theta$ can be seen in \fref{Lk-model} (d) - (f). Starting from the Lieb lattice ($\theta=\ang{0}$), there are Dirac points belonging to untilted Dirac cones at the four corners of the first Brillouin zone, the M points. When $\theta$ is increased, each of those Dirac points splits into four, two upper Dirac points which move along the M - K/K' direction and two lower Dirac points moving along the M - $\Gamma$ direction. This movement of Dirac points along high-symmetry directions away from the high-symmetry points is known to cause the Dirac cones to tilt in the respective direction of movement \cite{Yang2016, Goerbig2008}. When $\theta$ reaches \ang{30} (kagome lattice) the lower Dirac points merge into a parabolic band touching, and the remaining six Dirac cones at the corners K/K' of the first Brillouin zone are again untilted.

\section{Tilted Dirac cones}

\subsection{Classification of tilted Dirac cones}

To classify the Dirac cones of the Lieb-kagome lattice by their tilting, we adapt a model from \cite{Milicevic2019} to one dimension, considering a cut through a Dirac cone along its tilt direction. In our model, a Dirac cone is described by two intersecting lines with slopes of $s_1=v_0+v$ and $s_2=v_0-v$, respectively. Here the parameter $v$ describes the cones' opening angle (being wider for smaller $v$), while a non-zero $v_0$ introduces a tilt to the cones. Dirac cones can be classified by their tilt into four types (see \fref{DC-classification_1D}): Regular (untilted) type I cones, weakly tilted type I cones, strongly tilted type II cones, and critically tilted type III cones, which form the threshold between type I and II. In our model, this classification can be performed using an effective tilt parameter $v_0/v$ as shown in \fref{DC-classification_1D}.


\begin{figure}[htbp]
\includegraphics[width=0.95\columnwidth]{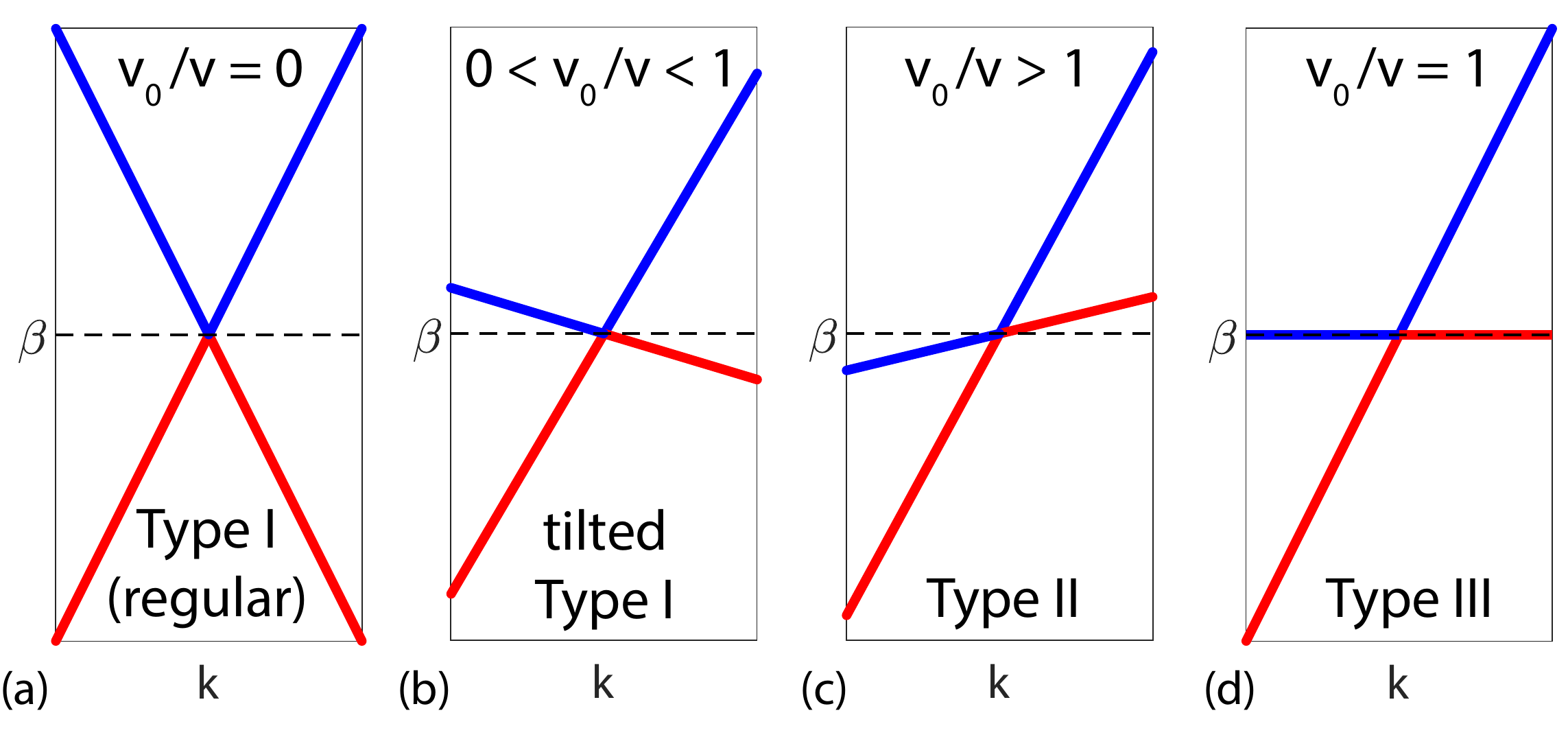}
\caption{\label{fig:DC-classification_1D} Classification of tilted Dirac cones in 1D by projecting them onto their tilt direction, which results in two intersecting lines. From the slopes of the lines, an effective tilt parameter $v/v_0$ can be calculated and used to classify Dirac cones into types I, II, and III. Based on \cite{Milicevic2019}.}
\end{figure}

Parameters $v$ and $v_0$ can be calculated from a 1D band structure such as \fref{Lk-model} (c) by fitting its data with two lines in the vicinity of the Dirac point. With this the degree of tilting and classification of Dirac cones can be tracked while varying different parameters in the band structure calculations. 

\subsection{Evolution of tilted Dirac cones in the Lieb-kagome lattice}

\fref{v0v-full} (a) shows $v_0$, corresponding to the degree of tilting, and $v_0/v$, the parameter used for classification, over the shearing angle $\theta$ for the upper and lower Dirac cones of the Lieb-kagome lattice. The edge cases of $\theta = \ang{0}$ and $\theta = \ang{30}$, as well as values of $\theta$ very close to them, were excluded. For both edge cases, there is only one type of Dirac cone with $v_0=v_0/v=0$ (i.e., no tilt), and for $\theta$ very close to those limits, there are not enough data points to perform reliable fits because the Dirac points are too close together. 

The degree of tilting ($v_0$) of both types of Dirac cones is the highest and almost identical, for small $\theta$. When $\theta$ increases, the tilt of the upper Dirac cones decreases almost linearly, while that of the lower Dirac cones resembles a curve of the form $y(x)=a-b\cdot \exp{(c\cdot x)}$, staying almost constant for low $\theta$ and rapidly decreasing for high $\theta$, until the tilt of both types of Dirac cones vanishes as $\theta$ approaches \ang{30}. For $\theta$, aside from the edge cases, the lower Dirac cones are therefore always tilted more strongly than the upper cones. 

The effective tilt parameter $v_0/v$ approaches 1 for both upper and lower Dirac cones for low $\theta$, meaning that they both approach the border case of critically tilted (type III) Dirac cones. For the upper Dirac cones, $v_0/v$ decreases with rising $\theta$, making them weakly tilted (type I) cones, while it increases for the lower Dirac cones, meaning they become strongly tilted (type II) cones. 

Another parameter that can be used to influence the Dirac cones' tilt is $n_{exp}$, the parameter describing how quickly next-nearest neighbor coupling decays with the distance between lattice sites (a higher $n_{exp}$ means that coupling decreases more quickly). This could be influenced in experiments by tailoring the coupling between lattice elements, for example, in photonic lattices through the choice of parameters for writing waveguides or the excitation wavelength \cite{el2019corner}. We therefore also calculated $v_0$ and $v_0/v$ in dependence on $n_{exp}$ with a fixed $\theta$ of \ang{15}, the results of which are shown in \fref{v0v-full} (b). 

Starting again with the degree of tilting, the lower Dirac cones show a stronger tilt for all values of $n_{exp}$. The difference between the two types of cone is the largest for small $n_{exp}$. With rising $n_{exp}$, $v_0$ initially increases for the upper Dirac cones and decreases for the lower Dirac cones, until both reach mostly constant values, which happens at a lower value of $n_{exp}$ for the lower cones. 

The difference in the effective tilt parameter of the upper and lower Dirac cones is also most pronounced for low $n_{exp}$, where again the upper Dirac cones are type I and the lower Dirac cones are type II. When $n_{exp}$ increases, both the upper and lower Dirac cones approach type III cones, which again happens at a lower \nexp~for the lower Dirac cones.

\begin{figure}[htbp]
\includegraphics[width=\columnwidth]{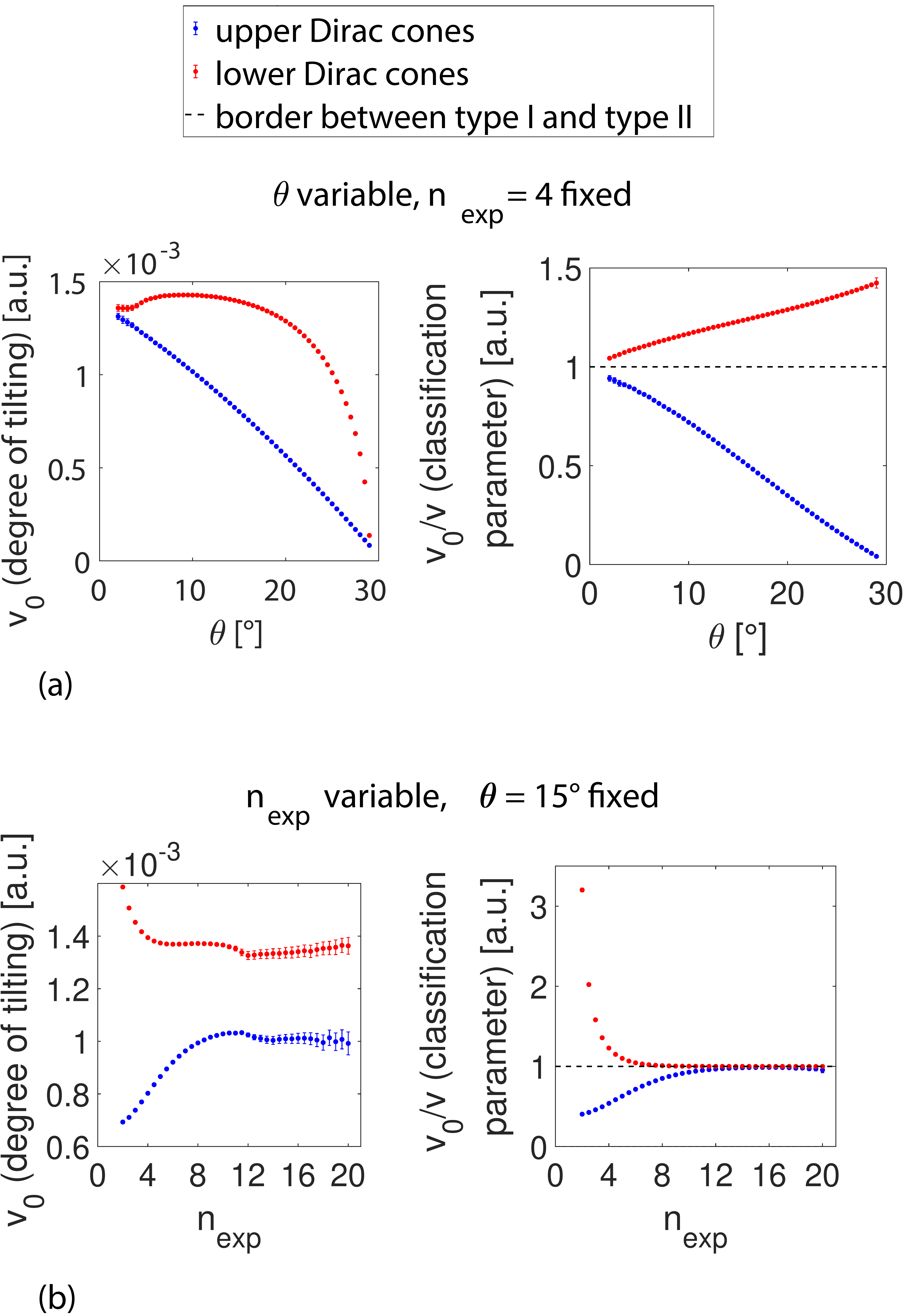}
\caption{\label{fig:v0v-full} Degree of tilting and classification of the upper and lower Dirac cones of the Lieb-kagome lattice \textbf{(a)} as a function of $\theta$, \textbf{(b)} as a function of \nexp. Error bars were calculated from the uncertainty of the fits performed to calculate these values.}
\end{figure}

\subsection{Discussion}

We have shown how the shearing angle $\theta$ and the parameter \nexp~can be used to influence the tilt of Dirac cones in the Lieb-kagome lattice, making it possible to tailor them to type I, II or III Dirac cones. For most parameters, the upper Dirac cones are type I, the lower Dirac cones are type II cones, but both approach critically tilted type III cones for low $\theta$ or high \nexp. In both cases, however, the Dirac cones are very close together in k-space, making it difficult to excite and study them individually. Based on our studies the best way to realize type III Dirac cones in the Lieb-kagome lattice is to increase \nexp, e.g.~by decreasing the refractive index increment of the waveguides or increasing the excitation wavelength, until the lower Dirac cones approach type III, which happens at a lower \nexp~than for the upper Dirac cones, so the Dirac cones are still well separated. 

\section{Asymmetric conical diffraction}

\subsection{Photonic lattices and conical diffraction}

Recently, both the Lieb and the kagome lattice have been realized as superlattices in electronic systems \cite{Slot2017, Li2018}, but for a general freely-tunable Lieb-kagome lattice such an approach would be unsuitable. A common way to study lattices with unconventional geometry is to use model systems. Particularly photonic lattices, which are periodic arrangements of weakly coupled waveguides, have been used to demonstrate a variety of phenomena known from solid-state physics. The evolution of light in such a photonic lattice can be described by the paraxial wave equation

\begin{equation}
\lambdabar \frac{\partial}{\partial z} \psi(x,y,z)-\left(\frac{\lambdabar^2}{2n_0}\nabla_\perp^2 + \Delta n(x,y,z)\right)\psi(x,y,z)=0,
\label{eq:paraxial}
\end{equation}

where $\psi(x,y,z)$ is the wave function, $\lambdabar=\lambda /2\pi$ is the reduced wavelength of light in the medium, $n_0$ is the refractive index of the medium, and $\Delta n(x,y,z)=n(x,y,z)-n_0$ is the refractive index increment between the waveguides and the medium. This equation is mathematically equivalent to a time-dependent two-dimensional Schrödinger equation, with the most striking difference being that time is replaced by the propagation distance $z$. This means that spatial light evolution in a photonic lattice is equivalent to time-evolution of the electron-wave function in a crystal lattice, and that the dispersion relation $\beta_n(k_x,k_y)$ of a photonic Lieb-kagome lattice can be described by the same tight-binding model we introduced in Sec.~\ref{sec:lattice}.

The presence of Dirac cones leads to a phenomenon called conical diffraction, where the lattice diffracts light into a ring with constant thickness and a radius which increases linearly with propagation distance. This was first demonstrated in simulations and experiments in 2007 by Peleg et al.~\cite{Peleg2007} using a honeycomb lattice. Asymmetric conical diffraction arising from tilted Dirac cones was demonstrated in simulations by Zhong et al. in 2019 \cite{Zhong2019a}, who observed rings moving transversally in the tilt direction of the cones during propagation 

\subsection{Beam propagation simulations of conical diffraction in the Lieb-kagome lattice}

In order to verify the existence of tilted Dirac cones in the Lieb-kagome lattice we studied conical diffraction in photonic lattices using a split-step beam propagation simulation. We chose the parameters of the simulation to be typical for photonic lattices produced by direct laser writing in fused silica. A summary of the most important parameters is provided in table \ref{tab:1}. \\

\begin{table}[h]
\begin{center}
\begin{tabular}{|M{0.75\columnwidth}|M{0.2\columnwidth}|} \hline
waveguide diameter (FWHM)& \mm{3} \\ \hline
maximal refractive index increment \newline of waveguides& \SI{1.3e-3}{} \\ \hline 
waveguide distance & \mm{20} \\ \hline 
excitation wavelength & 700\,nm \\ \hline
diameter of excitation beam (FWHM)& \mm{100} \\ \hline
propagation distance & 60\,mm \\ \hline
\end{tabular}
\end{center} 
\caption{Summary of the parameters of the beam propagation simulations.}
\label{tab:1}
\end{table}

In order to excite the Dirac cones of the lattice the positions of the Dirac points in k-space were calculated and they were targeted with a superposition of plane-waves with the corresponding phases. The resulting light field was then superimposed with a Gaussian beam in order to excite a finite region in k-space around the Dirac points and furthermore to leave enough space for the light to spread into the characteristic ring pattern. To observe full, unbroken and symmetric conical diffraction it is typically necessary to excite several Dirac cones at once \cite{Ablowitz2009}. For the Lieb and kagome lattices we therefore excited all Dirac points at the corners of the first Brillouin zone, four for the Lieb lattice and six for the kagome lattice (see \fref{Lk-model} (d) and (f)); for transition lattices with different $\theta$ we excited the six upper Dirac cones closest to the corners of the first Brillouin zone (see \fref{dp-tilt}). As an example, the excitation light field for the kagome lattice is shown in \fref{exc-comp} in real and Fourier space.

\begin{figure}[htbp]
\includegraphics[width=0.95\columnwidth]{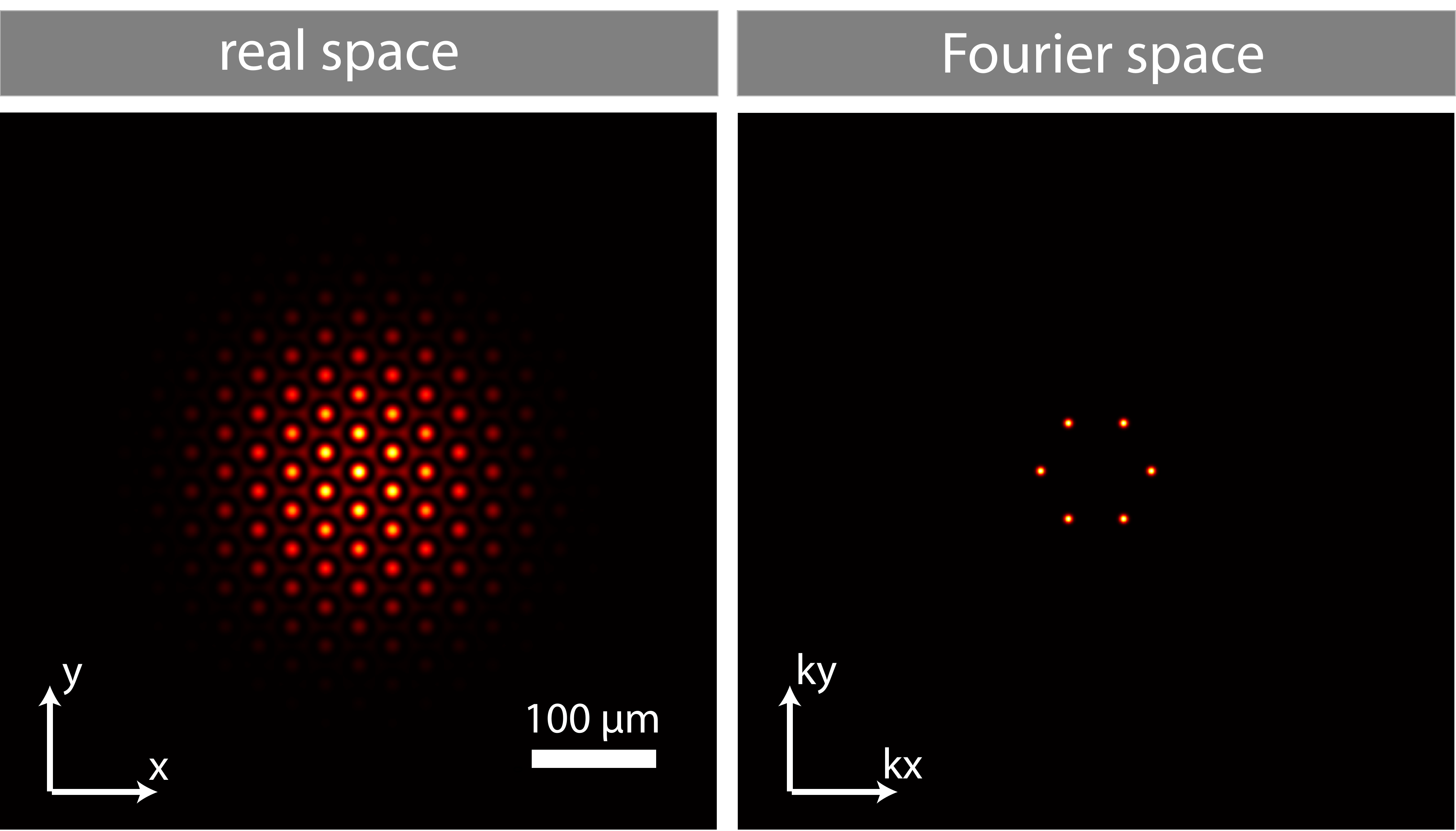}
\caption{\label{fig:exc-comp}Excitation light field for the kagome lattice in real and Fourier space, targeting six Dirac cones.}
\end{figure}

\begin{figure}[htbp]
\includegraphics[width=0.95\columnwidth]{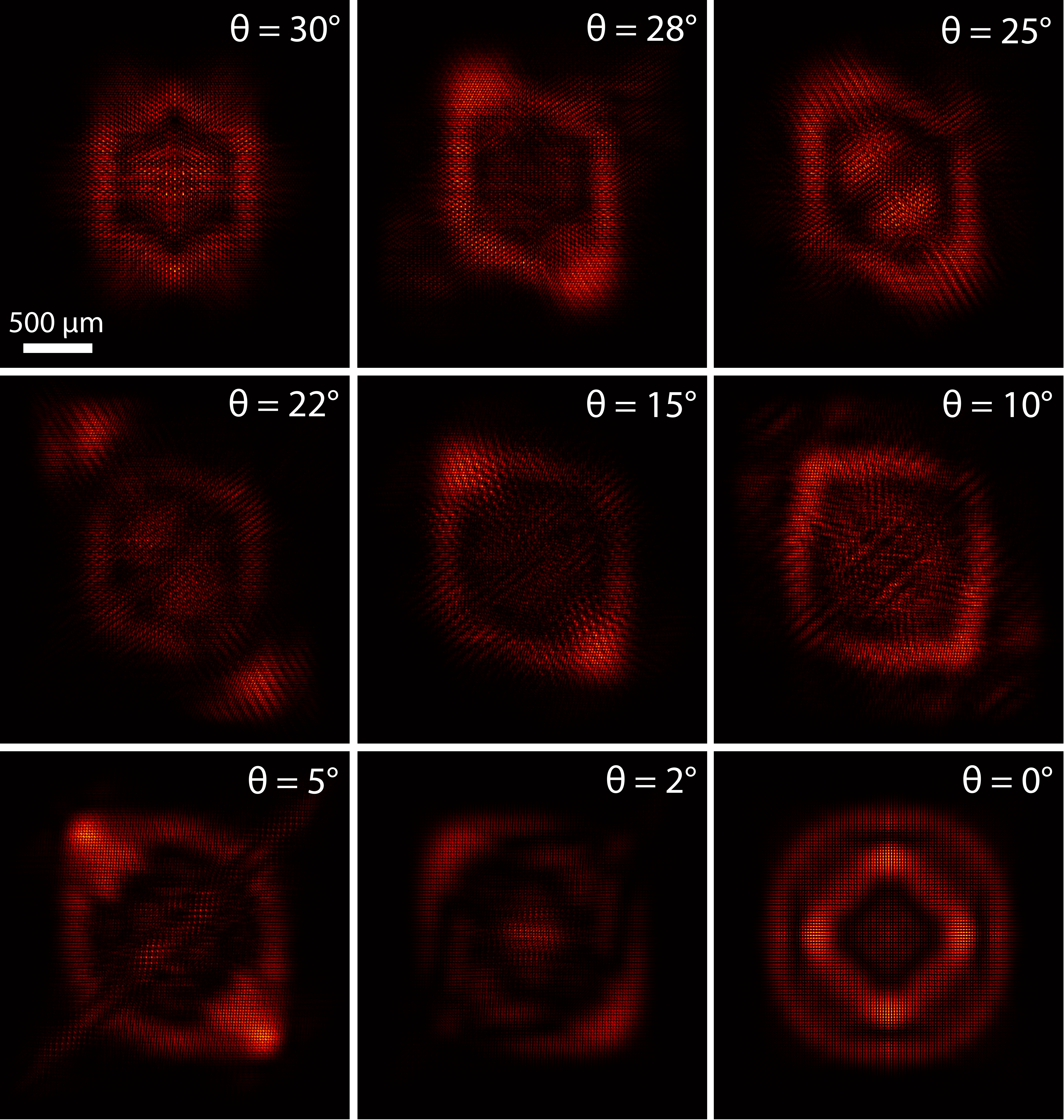}
\caption{\label{fig:ct1} Diffraction patterns after a propagation of 60\,mm through Lieb-kagome lattices with different shearing angles $\theta$, showing asymmetric conical diffraction for \ang{0} < $\theta$ < \ang{30}. For these patterns, selected Dirac points were excited by a superposition of plane waves. For the cases of regular Dirac cones the six ($\theta=\ang{30}$), respectively four ($\theta=\ang{0}$) Dirac points at the K/K' points were excited. For all other cases the six Dirac points closest to the K/K' points were excited.}
\end{figure}

\fref{ct1} shows the resulting diffraction patterns on the output facets of Lieb-kagome lattices with selected shearing angles $\theta$ after a propagation of 60\,mm, where the characteristic ring pattern of conical diffraction can be observed for all $\theta$. In the case of the kagome lattice ($\theta=\ang{30}$) and the Lieb lattice ($\theta=\ang{0}$), when the Dirac cones are untilted, the patterns are the most symmetric. In transition states (\ang{30} > $\theta$ > \ang{0}) the diffraction patterns become elliptic with varying eccentricities. The patterns are two-fold mirror symmetric with one symmetry axis in the tilt direction of the upper Dirac cones and the other in the tilt direction of the lower Dirac cones, which are perpendicular to each other. In all cases, some light remains in the middle of the ring, which is common with conical diffraction, and is usually attributed to the influence of bands other than those forming the Dirac cones \cite{Peleg2007}. For very small angles, in \fref{ct1} for $\theta=\ang{2}$, the ring pattern is broken: There first appear two separate half-circles moving apart in the tilt direction of the upper Dirac cones, followed by the same in the tilt direction of the lower Dirac cones. We attribute this to the fact that in this case the upper and lower Dirac cones are so close in k-space that our light field excites both at the same time. 

\begin{figure}[htbp]
	\centering
		\includegraphics[width=0.6666666666\columnwidth]{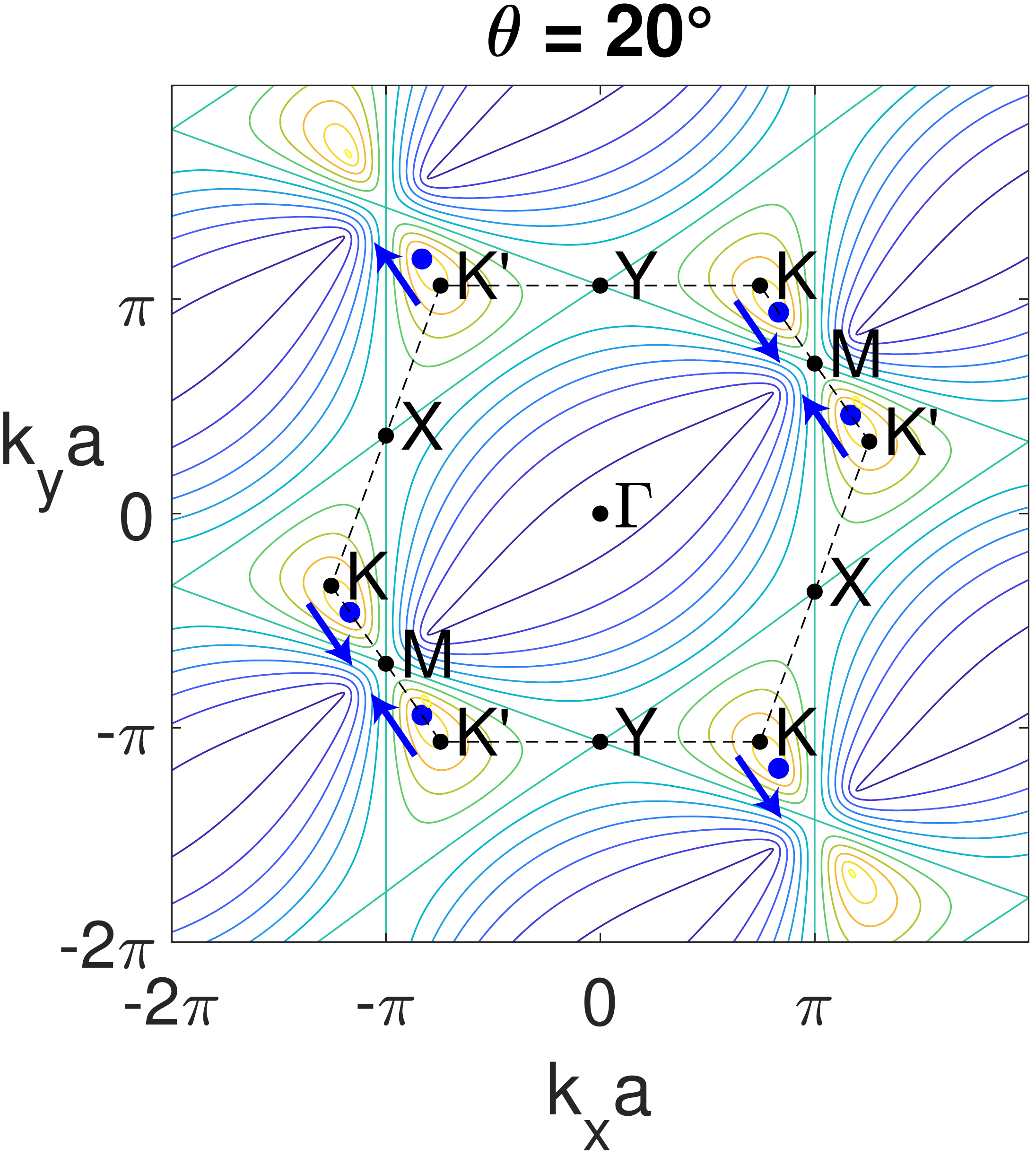} 
		\caption{k-space positions of the upper Dirac points (marked in blue) of Lieb-kagome lattices in transition states excited to generate conical diffraction. The arrows mark the tilt directions of the corresponding Dirac cones.}
	\label{fig:dp-tilt}
\end{figure}

When, starting from the Lieb lattice at $\theta=\ang{0}$, $\theta$ is increased, each Dirac point of the lattice splits into four, two upper Dirac points and two lower Dirac points, as seen in \fref{Lk-model} (d) - (f). In such a group of four Dirac points, the upper Dirac cones are tilted towards each other, while the lower Dirac cones are tilted away from each other. Of the six Dirac cones excited for the simulations in \fref{ct1} three are therefore tilted in one direction and the other three in the opposite direction, as shown in \fref{dp-tilt}. To show evidence of tilted Dirac cones in the Lieb-kagome lattice we need to specifically excite only Dirac cones tilted in one direction. The result of this approach is presented in \fref{60-k} for lattices with $\theta=$\ang{30}, \ang{25} and \ang{22}. In the left column all six Dirac cones shown in \fref{dp-tilt} were excited, in the middle column only those tilted diagonally upwards, and in the right column only those tilted diagonally downwards. The blue circles in \fref{60-k} mark the position and extent of the excitation light fields. When only three Dirac cones are excited, the symmetry of the diffraction patterns is reduced regardless of the shearing angle $\theta$. Nevertheless, in the case of the kagome lattice ($\theta=\ang{30}$) the patterns remain centered on the point where the lattice was excited, as is expected for regular conical diffraction. For $\theta=\ang{25}$ and $\theta=\ang{22}$, in contrast, we observe a clear shift of the diffraction patterns in the tilt direction of the Dirac cones when only cones tilted in one direction are excited. This behavior proves that the Dirac cones in question are indeed tilted \cite{Zhong2019a}.


\begin{figure}[htbp]
\includegraphics[width=\columnwidth]{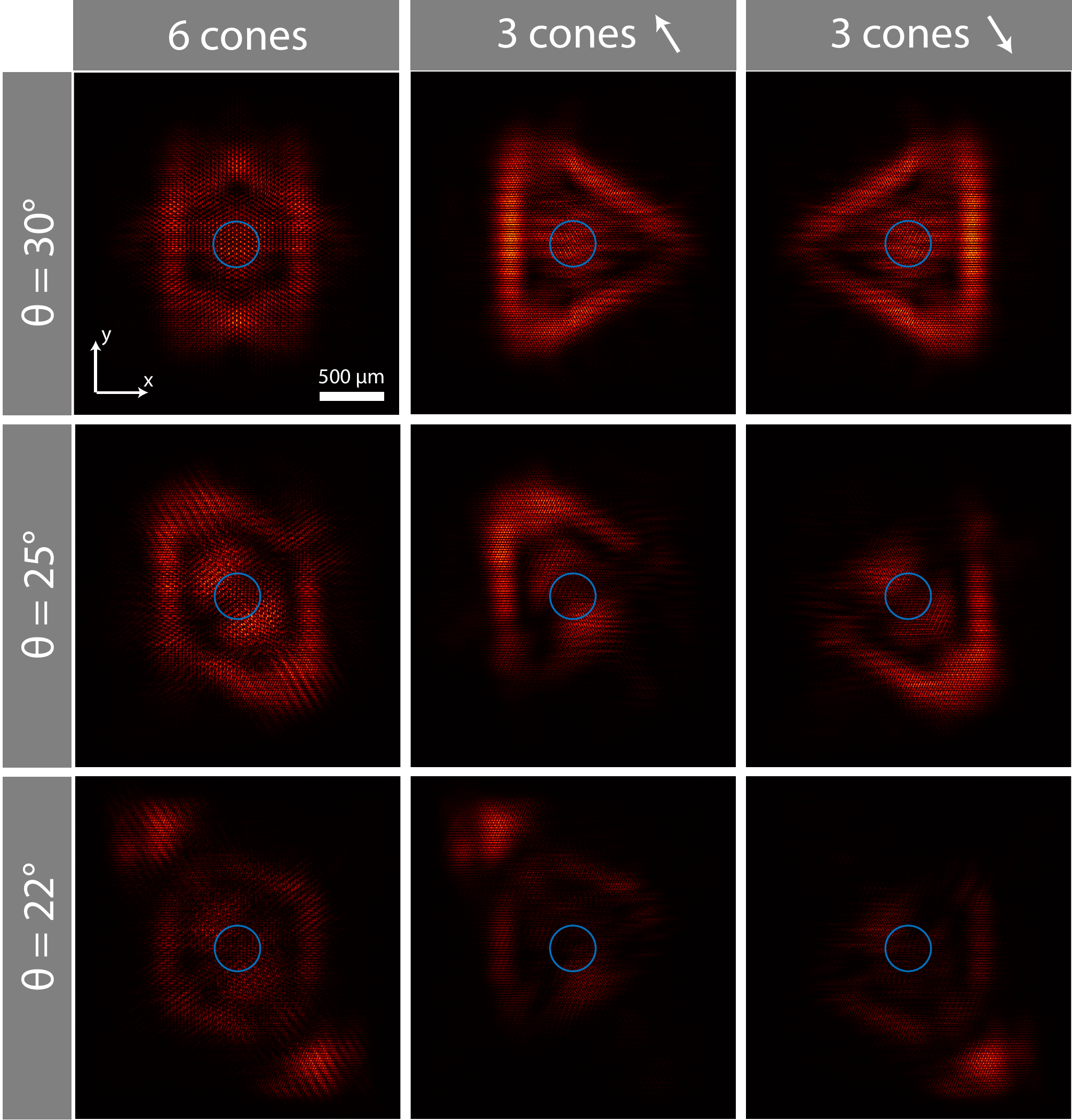}
\caption{\label{fig:60-k} 
Diffraction patterns after a propagation of 60\,mm through Lieb-kagome lattices with different shearing angles $\theta$, when different selections of Dirac cones are excited. The blue circles indicate the position and size of the excitation light fields.}
\end{figure}

\section{Experiments in fused silica}

In order to verify our results experimentally, we fabricated photonic lattices in fused silica (\ce{SiO2}) using the direct laser writing technique \cite{hanafi2022localizedAOM,hanafi2022APL}. Table \ref{tab:2} shows the main differences in parameters compared to the simulations presented so far. \fref{back} shows microscope images of the back facets of two of the lattices we fabricated, one with $\theta=\ang{30}$ (kagome lattice), consisting of 2415 waveguides, and one with $\theta=\ang{25}$, consisting of 1536 waveguides. 

\begin{table}[h]
\begin{center}
\begin{tabular}{|M{0.65\columnwidth}|M{0.3\columnwidth}|} \hline
waveguide dimensions (FWHM)& $ \mm{3}\times \mm{6}$ \\ \hline
waveguide distance & \mm{18} \\ \hline 
excitation wavelength & 532\,nm \\ \hline
propagation distance & 40\,mm \\ \hline
\end{tabular}
\end{center} 
\caption{Summary of the parameters of experiments and corresponding simulations.}
\label{tab:2}
\end{table}

\begin{figure}[htbp]
\includegraphics[width=0.8\columnwidth]{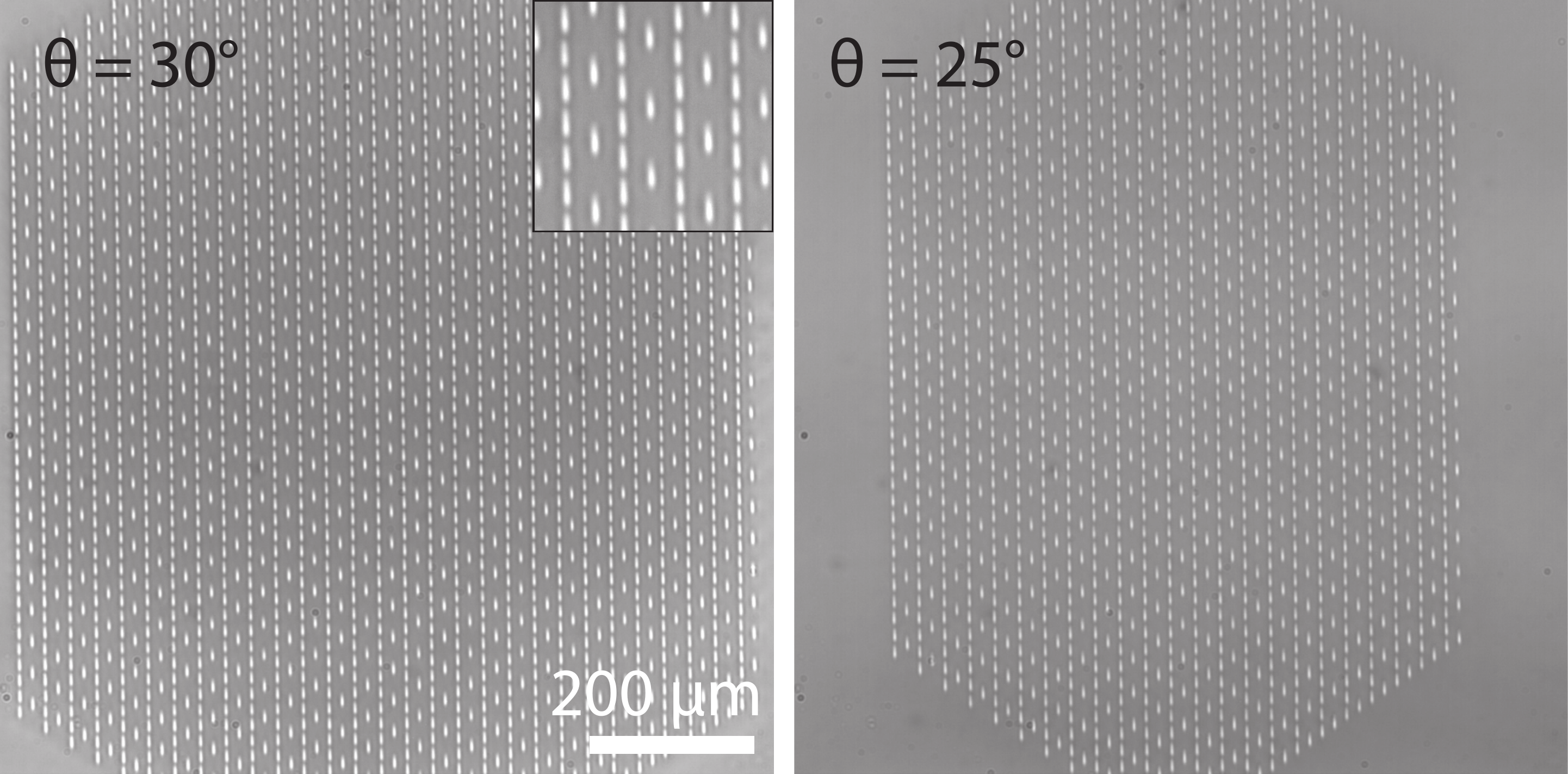}
\caption{Back facet of \ce{SiO2} sample with laser written waveguides, forming two photonic lattices. The inset is a close-up of some waveguides, showing their elliptical cross-sections.\label{fig:back} }
\end{figure}

Our direct laser writing setup uses a pulsed laser with a wavelength of 1030\,nm and a pulse length of approximately 250\,fs. Waveguides are written by translating the sample in a direction perpendicular to the laser beam (transversally) on a motion-controlled stage. A spatial light modulator (SLM) is used to counteract aberrations and compensate dependencies of the waveguide properties on the depth in the sample. Our setup for probing the resulting photonic lattices uses a continuous wave laser with a central wavelength of 532\,nm. An SLM is used to replicate the light fields used for excitation in the simulations, which were shown in \fref{exc-comp}. A more detailed description of both setups, including sketches, can be found in \cite{hanafi2022localizedAOM}.     

A known drawback of using a transversal writing scheme is that the waveguides fabricated this way exhibit elliptical instead of circular cross-sections, which can be seen in the inset of \fref{back}. This effect can currently only be partially compensated for, and known mitigation methods did not prove effective in our case. The elliptical waveguides introduce a small anisotropy in the coupling constants, which slightly deform the band structure of the resulting photonic lattices. Simulations with elliptical waveguides (\fref{ell}) show, that this causes conical diffraction to degenerate into a line pattern. In order to make patterns comparable for different values of the shearing angle $\theta$ we rotated the lattices by a $\theta$-dependent angle of $\varphi_{rot}=(90^{\circ}-\theta)/2+90^{\circ}$, which places the tilt direction of the upper Dirac cones in the y-direction and orients the line patterns along the same direction. 

\begin{figure}[htbp]
\includegraphics[width=0.95\columnwidth]{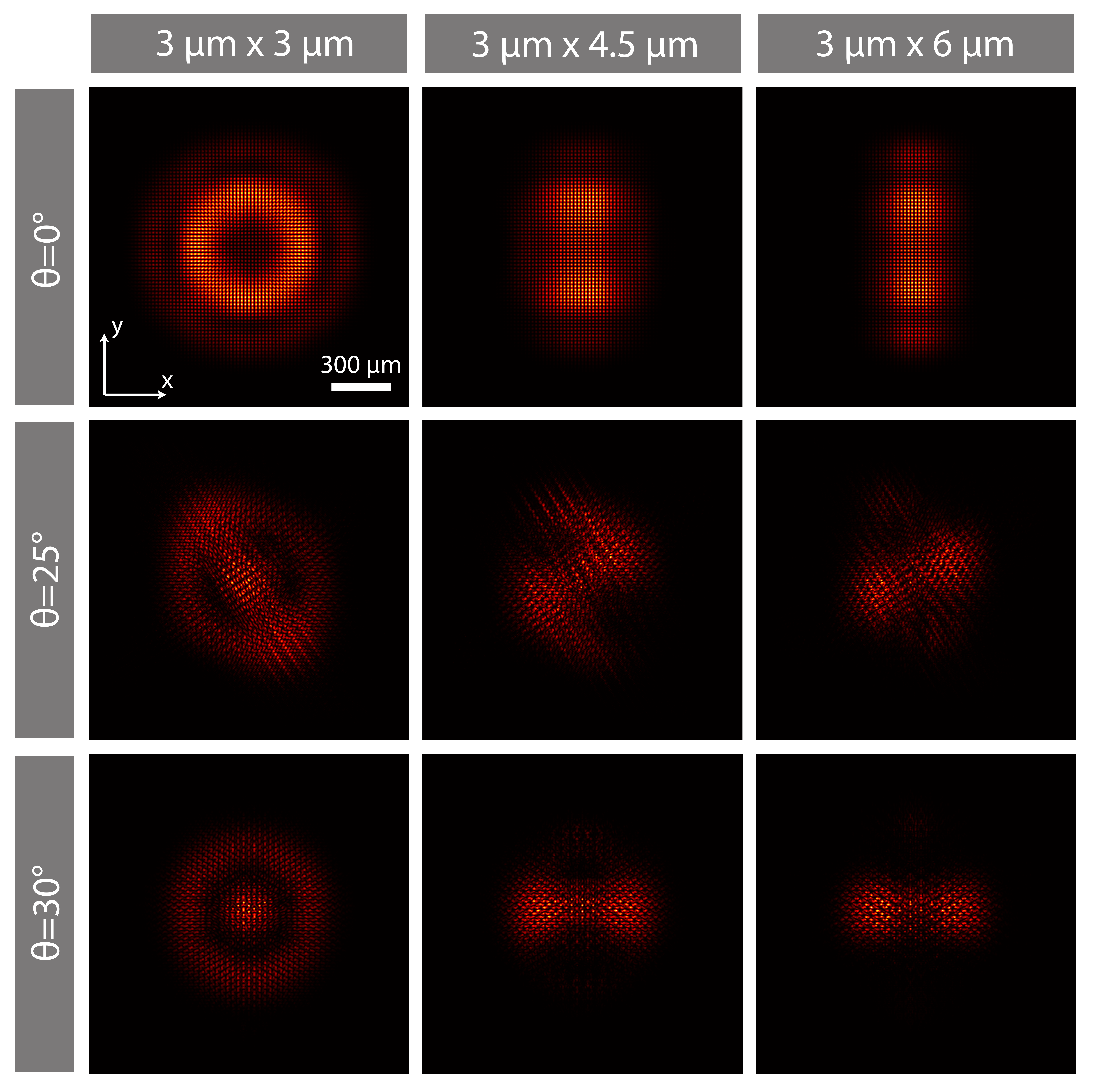}
\caption{Simulations demonstrating the deterioration of conical diffraction when the cross-section of waveguides changes from circular to elliptical.\label{fig:ell} }
\end{figure}

\fref{exp} shows the diffraction patterns after propagation of 40\,mm through the lattices described above, along with simulations using the parameters from Table \ref{tab:2} for comparison. To show the effect of tilting the same selections of Dirac cones as in \fref{60-k} were excited, which is again shown in the different columns. As discussed above, due to the elliptical cross-sections of the waveguides, the patterns are line-shaped with one or two bright spots rather than the circular patterns expected for conical diffraction. Nevertheless, in the simulations presented in \fref{exp}, the shift of the diffraction patterns' centers in the direction in which the Dirac cones are tilted can still be observed. This is the characteristic for tilted Dirac cones. The images obtained from experiments generally match the simulations, but exhibit more light remaining in the center of the pattern, i.e.~not diffracting. We attribute this to distortions of the band structure caused by the waveguides' elliptical cross-sections and other experimental inaccuracies.  

\begin{figure}[htbp]
\includegraphics[width=0.95\columnwidth]{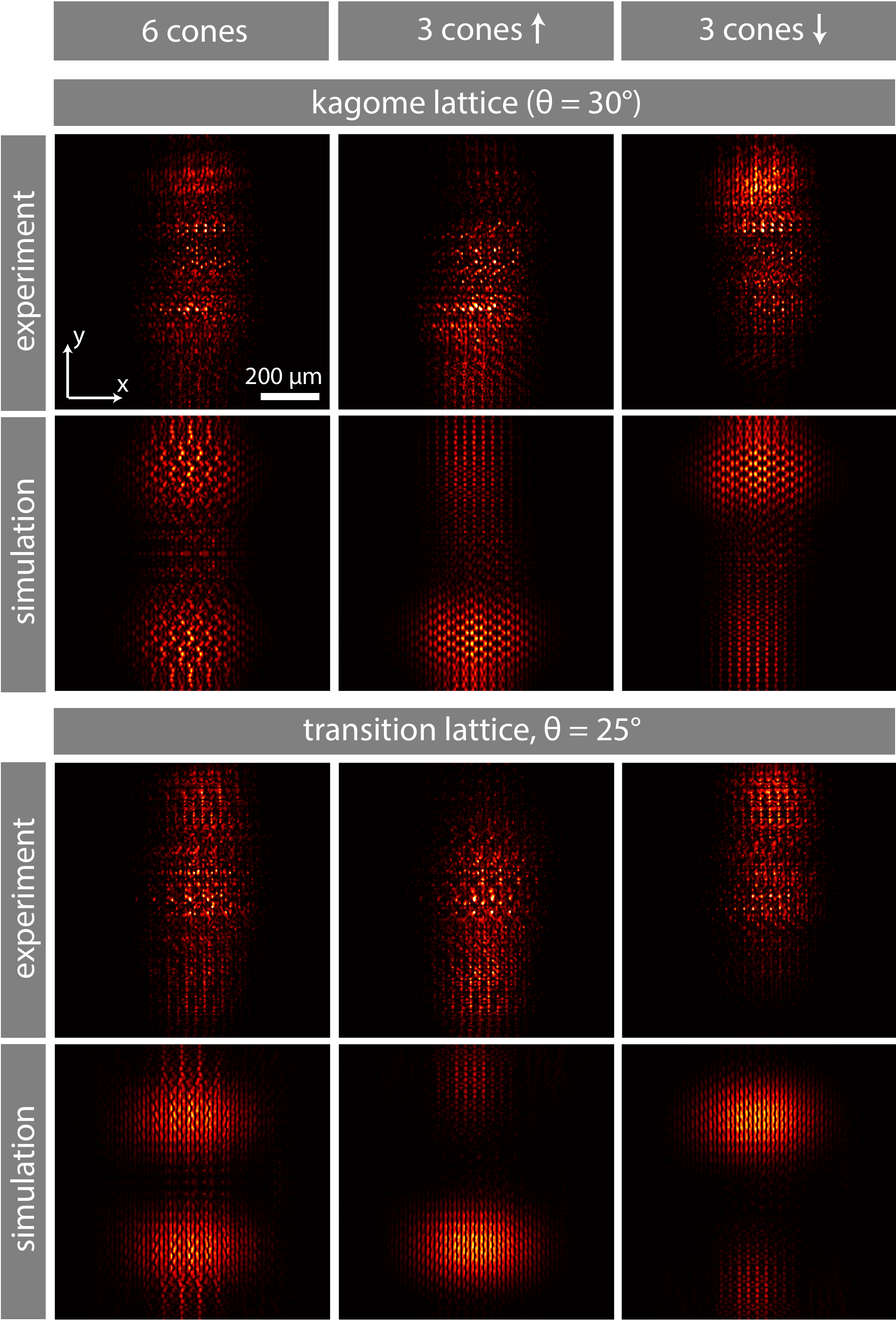}
\caption{Diffraction patterns after propagating 40\,mm through the lattices shown in \fref{back}, along with simulations for comparison. The same selections of Dirac cones as in \fref{60-k} were excited.\label{fig:exp} }
\end{figure}

\section{Conclusion}

In this work we have studied the tilted Dirac cones of the Lieb-kagome lattice in depth using the tight-binding method. We have shown how their tilting depends on the shearing angle $\theta$, which describes the transition between Lieb and kagome lattices and on the relative strength of next-nearest neighbor interaction, and how these parameters can be used to tune the Dirac cones of the lattice to type I, II or III. This provides future researchers with a way to deliberately engineer these different Dirac cones, and use them to study novel phenomena. After that, we studied realizations of Lieb-kagome lattices as photonic lattices using split-step beam propagation simulations. We have for the first time shown conical diffraction in transition states between Lieb and kagome lattice, where we found evidence of Dirac cones tilted in different directions. In a visionary approach we fabricated on a fused silica chip large-scale photonic lattices consisting of thousands of single-mode waveguides. Although experimental challenges exist in the form of elliptical waveguides instead of circular waveguides, we were able to prove the signature of tilted Dirac cones. We expect that the presented experiments can be further improved by switching to a longitudinal writing scheme or an alternative fabrication technique. Our studies further the understanding of the Lieb-kagome lattice and tilted Dirac cones in general and provide a basis for further research into these subjects.

\section{Author information}
Jean-Philippe Lang and Haissam Hanafi contributed equally to this work.

\clearpage




%

\end{document}